\documentclass{article}
\usepackage{spconf,amsmath,graphicx,hyperref}

\usepackage{cite}
\usepackage{amsmath,amssymb,amsfonts}
\usepackage{algorithmic}
\usepackage{graphicx}
\usepackage{textcomp}
\usepackage{multirow}
\usepackage{booktabs}
\usepackage{booktabs}
\usepackage[table]{xcolor}
\usepackage{placeins}
\usepackage{xcolor}
\usepackage{makecell}

\title{Spatial Audio Motion Understanding and Reasoning}
\name{Arvind Krishna Sridhar$^{ \dagger}$ \qquad Yinyi Guo$^{\dagger}$ \qquad Erik Visser$^{\dagger}$}
  
  \address{$^{\dagger}$ \textit{Qualcomm Technologies Inc.} \\
  San Diego, CA
      }
\begin{document}
\ninept
\maketitle
\begin{abstract}
Spatial audio reasoning enables machines to interpret auditory scenes by understanding events and their spatial attributes. In this work, we focus on spatial audio understanding with an emphasis on reasoning about moving sources. First, we introduce a spatial audio encoder that processes spatial audio to detect multiple overlapping events and estimate their spatial attributes, Direction of Arrival (DoA) and source distance, at the frame level. To generalize to unseen events, we incorporate an audio grounding model that aligns audio features with semantic audio class text embeddings via a cross-attention mechanism. Second, to answer complex queries about dynamic audio scenes involving moving sources,  we condition a large language model (LLM) on structured spatial attributes extracted by our model. Finally, we introduce a spatial audio motion understanding and reasoning benchmark dataset and demonstrate our framework's performance against the baseline model.
\end{abstract}
%
\section{Introduction}
\label{sec:intro}
Recent advances in audio question answering (AQA) have led to the development of state-of-the-art models capable of generating relevant answers from a given question and an audio clip. Building on this progress, we argue that spatial audio question answering (SAQA) represents the next frontier in context-aware audio understanding. SAQA extends AQA to multichannel audio, implicitly integrating several traditional audio understanding tasks, including audio event detection (AED), audio captioning, audio localization and detection\cite{shimada2025stereo}, source separation\cite{nugraha2016multichannel} and object tracking\cite{li2025patch}. Spatial AQA has applications in smart surveillance, robotics, autonomous vehicles, and augmented reality. To interpret complex auditory scenes in these domains, SAQA systems must recognize diverse audio events and reason about dynamic source movements. This work advances SAQA with a focus on movement reasoning, bridging the gap between static spatial representations and real-world dynamics. The DCASE challenge Sound Event Localization and Detection (SELD) task \cite{shimada2025stereo}, has significantly advanced research in spatial audio by promoting models that jointly estimate event classes and their spatial attributes. While these models have achieved strong performance on benchmark datasets, they are typically trained on a fixed set of audio events and predefined spatial configurations, limiting their adaptability to unseen classes. To overcome the limitations of existing SELD and spatial localization models—which require costly data collection and retraining to incorporate new classes—we introduce a flexible and generalizable spatial audio encoder by leveraging text-to-audio grounding model\cite{xu2024towards}. Text-to-audio grounding is the task of detecting audio events in audioclip based on natural language \cite{yasuda2022echo}. From here on, we refer to this model as Audio Grounding Model (AGM) for brevity. Similar to CLAP\cite{yuan2024tclaptemporalenhancedcontrastivelanguageaudio}, ELSA \cite{spatially-aware} consists of a spatial audio encoder and a text encoder that learns a joint spatial attributes and semantic embedding via contrastive learning.  Although the spatial audio representation learned by ELSA is well-suited for tasks like semantic retrieval and spatial audio captioning, ELSA does not perform explicit question answering and lacks fine-grained temporal reasoning.

BAT \cite{zheng2024bat} introduced the first end-to-end SAQA model, combining a spatial audio encoder with a large language model (LLM) to process binaural audio and project it into a text-aligned representation for question answering. However, BAT faces key limitations: it depends on large-scale spatial QA pair generation and supports only static sources.  These constraints, coupled with the high computational cost of end-to-end multimodal training, hinder scalability and generalization to complex real-world environments. While end-to-end training can be parameter-efficient, it requires extensive data preparation and specialized training strategies such as curriculum learning\cite{sridharenhancing}. Moreover, fine-tuning such models to acquire new spatial reasoning skills often leads to catastrophic forgetting\cite{chen2022continual}. Motivated by these challenges, we propose an end-to-end training-free SAQA framework (Fig. \ref{fig:Spatial AQA framework}).

\noindent To the best of our knowledge, we are the first to make the following contributions:
\begin{itemize}
    \item \textbf{Dynamic Spatial Audio Spectrogram Transformer with Audio Grounding:} We design a Dynamic Spatial Audio Spectrogram Transformer (DSAST) model that detects multiple overlapping audio events and estimates their spatial attributes, direction of arrival (DoA) and source distance, at the frame level. To generalize and handle unseen audio events, we incorporate mono audio grounding, which aligns semantic audio class text embeddings with spatial audio representations through a cross-attention mechanism.
    \item \textbf{Spatial Audio Reasoning Framework:} We integrate the proposed DSAST model with a large language model (LLM) to enable spatial reasoning and answer complex audio queries.
    \item \textbf{Spatial Audio Motion Understanding and Reasoning Benchmark:} We introduce a benchmark focused on reasoning about source movements and evaluate our approach against state-of-the-art spatial AQA systems.
\end{itemize}

\section{Methodology}
\label{sec:methodology}
\begin{figure*}[t!]
\centering
\begin{minipage}[b]{1\linewidth}
  \centering
  \centerline{
  \includegraphics[width=1\textwidth,scale=1]{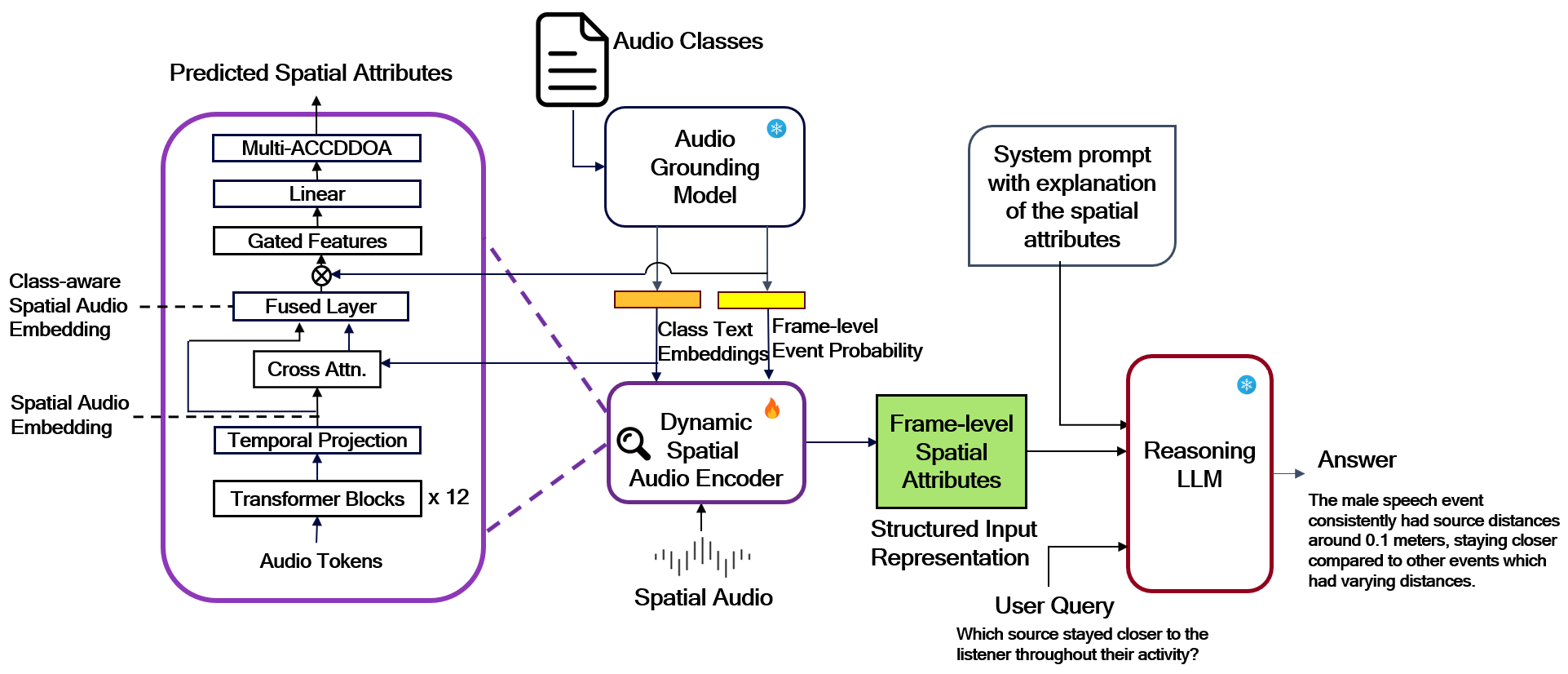}
  }
\end{minipage}
\caption{Spatial Audio Reasoning Framework}
\label{fig:Spatial AQA framework}
\end{figure*}

Our proposed spatial audio reasoning framework, as shown in Figure \ref{fig:Spatial AQA framework}, comprises of a spatial audio encoder and a reasoning LLM. The methodology explains in detail about the architecture of our DSAST with AGM model, spatial audio reasoning framework and the creation of spatial audio motion understanding and reasoning benchmark.
\subsection{Dynamic Spatial Audio Spectrogram Transformer with Audio Grounding}
\label{sec:Spatialaudiowgrounding}
The DSAST takes stereo audio as input and predicts, at frame level, the audio events along with their spatial attributes, direction of arrival (DoA) and source distance. We adopt the audio waveforms preprocessing steps from BAT\cite{zheng2024bat}. First, we compute the mel-spectrogram and interaural phase difference (IPD) from the waveforms. The two-channel mel-spectrogram is then combined with the sine and cosine of the IPD, providing a unified representation that captures both spectral content and spatial cues \ref{Equation:IPD}. These features are downsampled to a single channel, normalized, and passed through a GELU activation layer. Next, we apply patch embedding to convert the features into audio tokens, which are processed by a 12-layer Transformer encoder that learns spatial audio representations via self-attention. Next, we apply a temporal linear projection to capture variations in audio events over time.
\begin{equation}
\begin{split}
\mathbf{Z} &= [S_1; S_2; \cos(\text{IPD}) \times \text{melW}; \sin(\text{IPD}) \times \text{melW}], \\
\mathbf{Z} &\in \mathbb{R}^{4 \times T \times M}
\end{split}
\label{Equation:IPD}
\end{equation}
\vspace{-1mm}
where $T$: number of time frames, 
$M$: number of mel-frequency bins, 
$S_1, S_2 \in \mathbb{R}^{T \times M}$: per-channel spectral features for channels 1 and 2, 
$\mathrm{IPD} \in \mathbb{R}^{T \times M}$: interaural phase difference at each time–mel bin, 
$\mathrm{melW}$: mel filterbank

\begin{equation}
\begin{split}
(P, \mathbf{E}) &= \mathrm{AGM}(A, \mathcal{C}),\\
P &\in \mathbb{R}^{T \times N}, \quad \mathbf{E} \in \mathbb{R}^{N \times d},\\
\end{split}
\label{Equation:AGM}
\end{equation}
where $N$: number of classes, $A$: audio clip,
$\mathcal{C}$: list/set of audio class names,
$P$: frame-level class probability matrix ($T \times C$),
$d{=}512$: embedding dimensionality,
$\mathbf{E}$: class text embedding tensor ($N=50 \times d=512$).

To incorporate semantic audio class information into audio tokens, we utilize the Audio Grounding Model (AGM) which is trained for text-to-audio grounding (TAG) task. We provide the list of audio classes to detect as natural langauge queries and the audioclip to AGM and receive the frame-level class probabilities and audio class text embeddings as outputs, as shown in Equation \ref{Equation:AGM}. We fuse the audio tokens with class text embeddings using a cross-attention mechanism (Equation \ref{eq:crossattn_fuse}), aligning audio and text representations. The attention output is concatenated with the encoded audio features, while frame-level probabilities act as a gating mechanism to weight features by event confidence from the AGM. This design leverages the audio event detection confidence from AGM which improved the generalization of DSAST. Finally, the gated features pass through linear layers with \texttt{Tanh} and \texttt{ReLU} activations to predict DoA and source distance for each time frame. We use the Multi-ACCDDOA loss\cite{Krause2024SoundED} with MSE loss to train the spatial audio encoder while keeping the AGM frozen.

\begin{equation}
\begin{split}
\mathbf{H} &= \mathrm{Attn}\!\big(\mathbf{Z}_{\mathrm{enc}},\, \mathbf{E},\, \mathbf{E}\big)
\in \mathbb{R}^{T \times N \times d},\\
\mathbf{F} &= \big[\,\mathbf{Z}_{\mathrm{enc}};\, \mathbf{H}\,\big] \in \mathbb{R}^{T \times N \times 2d}\\
\widehat{\mathbf{F}} &= \mathbf{P}_{\mathrm{norm}} \odot \mathbf{F} \in \mathbb{R}^{T \times N \times 2d}
\end{split}
\label{eq:crossattn_fuse}
\end{equation}
where $\mathbf{Z}_{\mathrm{enc}} \in \mathbb{R}^{T \times N \times d}$: encoded audio token embeddings, 
$\mathbf{P}_{\mathrm{norm}} \in \mathbb{R}^{T \times N}$: normalized frame‑level class probabilities,
$\odot$: elementwise (Hadamard) product, broadcast along the feature dimension.

\subsection{Spatial Audio Reasoning Framework}
\label{sec:Spatialaudioreasoningframework}
The spatial audio reasoning framework, illustrated in Figure \ref{fig:Spatial AQA framework}, is a training-free approach that combines the DSAST model (Section \ref{sec:Spatialaudiowgrounding}) with structured prompting of the LLM using the predicted spatial attributes. The LLM receives three inputs: a system prompt, the structured spatial attributes, and the user question, and performs spatial reasoning to generate an answer. Spatial attributes are provided in an event-centric format as a list of JSON objects:
\textit{[ { 'Event': 'audio event 1', 'DoA': \textless list of n values\textgreater, 'Source distance': \textless list of n values\textgreater, 'Time frames': \textless list of n active times\textgreater }, \dots ]}. The system prompt defines the meaning of spatial attributes (DoA and source distance) and specifies the expected input and output formats. To enable reasoning across time frames for queries such as “How did the source distance change over time?”, we employ LLMs trained with chain-of-thought or reasoning-oriented prompting strategies.

\vspace{-1mm}
\setlength{\tabcolsep}{4pt} 
\begin{table*}[ht]
\centering
\caption{Question categories in spatial audio motion understanding and reasoning benchmark}
\begin{tabular}{p{3cm}p{2cm}p{10cm}p{2cm}}
\toprule
\textbf{Category} & \textbf{Question type} & \textbf{Example question} & \textbf{Answer} \\
\midrule
DoA \& its trajectory & Boolean & Did the speaker move noticeably from the left side toward the front over the course of the clip? & Yes \\
Distance Dynamics & MCQ & By the end of the male voice activity, was the speaker closer, farther, or about the same distance compared to the start? & About the same \\
Cross-Source Comparisons & Boolean & Did the laughter change its direction more than the male speaker did (i.e., show a larger left/right shift)? & No \\
Distance vs DoA  & MCQ & Which aspect changed more over the clip: the direction the voice came from or the distance to the listener? & Distance changed more \\
\bottomrule
\end{tabular}
\vspace{-3mm}
\label{Table:Questions}
\end{table*}

\subsection{Spatial Audio Motion Understanding and Reasoning Benchmark}
We propose a question-answering benchmark designed to evaluate spatial audio QA systems on their ability to reason about moving sound sources. To construct the benchmark data, we prompt GPT-5 mini with metadata from the test set of the DCASE 2025 Task 3 SELD challenge to generate QA pairs.
In the underlying audio scenes, the Direction of Arrival (DoA) ranges from \(-90^\circ\) (left) to \(+90^\circ\) (right), with \(0^\circ\) corresponding to the front. Source distance varies between \(0\,\text{m}\) and \(6\,\text{m}\) from the listener. While prior work such as BAT focuses on QA pairs that require direct estimation of DoA and distance, real-world queries often involve understanding spatio-temporal audio event relationships rather than purely numerical. To reflect this, we discretize DoA into five directional buckets: \textit{front-left} (\(+45^\circ\) to \(+90^\circ\)), \textit{front} (\(-15^\circ\) to \(+15^\circ\)), \textit{front-right} (\(-90^\circ\) to \(-45^\circ\)), \textit{slightly left} (\(+15^\circ\) to \(+45^\circ\)), and \textit{slightly right} (\(-45^\circ\) to \(-15^\circ\)).
To account for minor variations in predicted spatial attributes, we define a tolerance threshold: a change of at least \(5^\circ\) in DoA or \(0.005\,\text{m}\) in distance between frames is considered significant for movement reasoning. This design encourages models to reason about spatial dynamics rather than memorize exact numeric values.

We focus on generating questions across four categories, as detailed in Table~\ref{Table:Questions}. The first category, \textit{DoA \& its trajectory}, evaluates whether the model can track how the Direction of Arrival (DoA) of an audio source changes over time. The second category, \textit{distance dynamics}, tests the model's ability to reason about variations in source distance, e.g., ``Did the male speaker move closer to the listener?'' The third category, \textit{cross-source comparison}, involves comparing spatial attributes across two or more sources, such as ``Which audio source came closest to the listener during the clip?'' This requires identifying the minimum source distance across time frames and comparing them across sources. The fourth category, \textit{distance vs. DoA}, compares the rate of change of both spatial attributes. For example: ``Which aspect changed more over the clip: the direction or the distance from the listener?'' This requires computing attribute change rates and interpreting significance based on their ranges. For instance, a 1\,m change in distance (range: 0--6\,m) is more perceptually significant than a $10^\circ$ change in DoA (range: $-90^\circ$ to $+90^\circ$). For ease of evaluation, all QA pairs are formulated as either Boolean or single-answer multiple-choice (MCQ) questions. The full prompt format for generating this benchmark using GPT-5 mini is as follows: Explanation of the metadata format as described in Section \ref{sec:Spatialaudioreasoningframework}, rules to discretize DoA and tolerance threshold as defined in the previous paragraph, definitions of the question categories along with the examples as shown in Table \ref{Table:Questions}, output format definition for generating MCQ and boolean type questions Return a JSON array where each element has:
\{
  "clip\_id": "unique short ID",
  "type": "mcq-single | boolean",
  "prompt": "the question text",
  "choices": optional array for MCQ",
  "answer": "correct answer",
  "scoring": "choice\_match or exact"
\}

\begin{table}[t]
\centering
\caption{A comparison of spatial audio encoder performance. Unseen refers to the out-of-training domain audio classes dataset described in \ref{sec:datasets}. Columns: F-Score (in \%), DoA Error (in degrees), and Relative Distance Error. Here Dcase baseline refers to DCASE 2025 Task 3 baseline.}
\label{tab:spatial_encoder_eval}
\begin{tabular}{lccc}
\toprule
\makecell{\textbf{Spatial}\\\textbf{Audio Encoder}} & \textbf{F-Score} & \textbf{DoA Error} & \textbf{Rel. Dist. Error} \\
\midrule
DCASE baseline                & 0.260 & 23.0  & 0.332 \\ \hline
DSAST (Ours)                 & 0.225 & 24.39 & 0.305 \\ 
DSAST w/ AGM (Ours)          & 0.188 & 27.13 & 0.336 \\
\makecell{DSAST w/ AGM\\(Ours) (Unseen)} & 0.327 & 25.05 & 0.510 \\
\bottomrule
\vspace{-4mm}
\end{tabular}
\label{Table:spatialaudioencoder}
\end{table}

\begin{table*}[t]
\centering
\caption{SAQA performance on Spatial Audio Motion Understanding and Reasoning Benchmark Columns: 
DoA = Direction of Arrival \& its trajectory; DistDyn = Distance Dynamics; 
CrossSrc = Cross-Source comparisons; Dist vs DoA = Distance vs DoA; Overall = all questions. The questions are framed as MCQ-single answer or boolean type.}
\label{tab:cap5_models_rows}
\begin{tabular}{lcccccc}
\toprule
\textbf{Model} &
\textbf{DoA} &
\textbf{DistDyn} &
\textbf{CrossSrc} &
\textbf{Dist vs DoA} &
\textbf{Overall} \\
\midrule
BAT (baseline)         & 11.3\% & 3.8\%  & 20.0\% & 0.0\%   & 8.8\% \\ \hline
DSAST+Qwen7B (Ours)    & \textbf{35.8\%} & 26.9\% & 20.0\% & 0.0\%  & 20.7\% \\
DSASTw/AGM+Qwen7B (Ours)        & 26.4\% & \textbf{34.6\%} & \textbf{30.0\%}  & \textbf{33.3}\% & \textbf{31.1\%} \\
\bottomrule
\end{tabular}
\label{Table:Maintable}
\end{table*}

\vspace{-3mm}
\section{Experiments}
\subsection{Datasets}
\vspace{-1mm}
\label{sec:datasets}
 We train the spatial audio encoder models with audio only training set of STARS23 dataset \cite{Shimada2023starss23}. The stereo version of STARS23 comprises of 30k five-second audio recordings and 13 audio event classes. Important fact to note is that the DCASE 2025 Task 3 baseline is mentioned to have been fine-tuned with additional synthetic samples while we train our models only with the publicly available train set of STARS23 for reproducibility\cite{shimada2025stereo}. Evaluation is performed on the STARSS23 test set. To assess generalization, we built our 9 unseen isolated sound events dataset using dry source recordings from FOA-MEIR dataset\cite{yasuda2022echo}. We spatialize the audioclips using Spatial Scaper~\cite{roman2024spatial}, generating 3k five-second stereo clips with up to three overlapping events per frame. FOA outputs are converted to stereo following~\cite{shimada2025stereo}. We adopt the multi-ACCDDOA loss, evaluation metrics-F-score, DoA error, and Relative Distance Error, as well as feature extraction hyperparameters from~\cite{shimada2025stereo}. We generate the spatial audio evaluation benchmark using 1000 randomly chosen audio clips from the test set of STARS23 dataset using GPT-5 mini.
  
\subsection{Experiment Setup}
Our Audio Grounding Model (AGM) is trained following the phrase-level WSTAG framework described in \cite{xu2024towards}, utilizing the AudioCaps\cite{kim-etal-2019-audiocaps} dataset with extracted phrases from captions. AGM adopts the same hyperparameters and model architecture as \cite{xu2024towards}, comprising an audio encoder based on a CRNN with eight convolutional layers followed by a bidirectional GRU (BiGRU), and a text encoder consisting of a single word embedding layer with mean pooling. Frame-wise scores for each audio tag are computed by measuring the cosine similarity between frame-level audio embeddings and tag embeddings, followed by a sigmoid activation. The AGM remains frozen during DSAST model training. We consider the DCASE 2025 Task 3 baseline\cite{shimada2025stereo} as reference point for the Table \ref{Table:spatialaudioencoder}.

We initialize the 12 transformer layers of DSAST model with pretrained weights of Audio-MAE\cite{huang2022masked}. Both DSAST and DSAST w/ AGM are trained for 100 epochs with 5 warm-up epochs, a maximum learning rate of \(1\times10^{-3}\), batch size of 64 and layer-wise decay of 0.0005. Time and frequency masking are disabled to preserve temporal and spectral relationships. For the spatial audio reasoning framework, we use the DeepSeek-R1 distilled Qwen-7B model, hereafter called as Qwen-7B, from HuggingFace\cite{DBLP:journals/corr/abs-1910-03771} with greedy decoding asn the reasoning LLM. We consider BAT\cite{zheng2024bat}, which employs LLaMA-2 (7B), as the baseline for end-to-end question answering.

\begin{figure}[t!]
\begin{minipage}[b]{0.9\linewidth}
  \centering
  \centerline{
  \includegraphics[width=1.0\textwidth,scale=0.8]{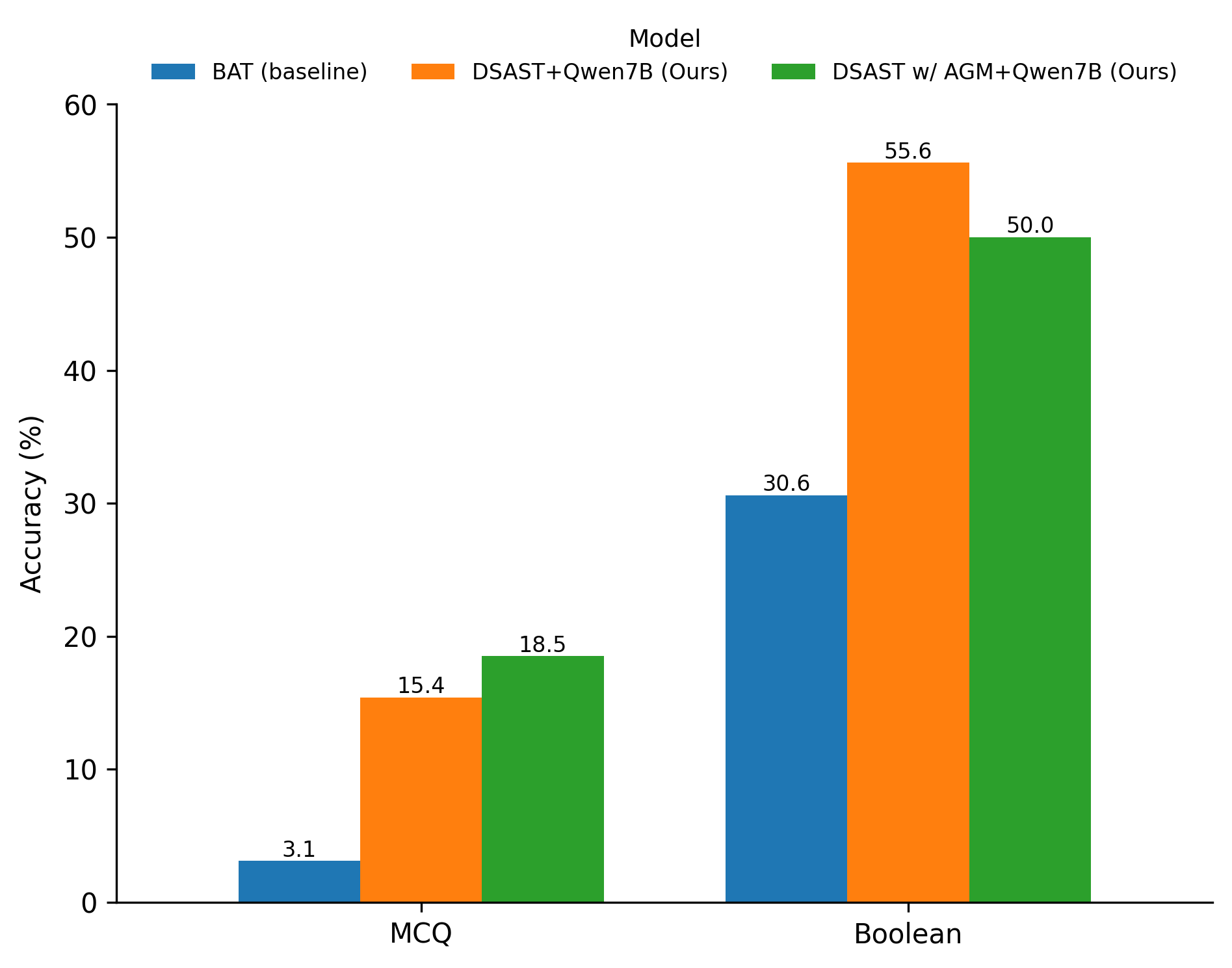}
  }
\end{minipage}
\caption{Barplot for SAQA performance by question type (MCQ vs Boolean). Blue: BAT (baseline), Orange: DSAST+Qwen7B (Ours), Green: DSAST w/ AGM+Qwen7B (Ours)}
\label{fig:barplotquestiontype}
\vspace{-3mm}
\end{figure}

\section{Results}
The Table \ref{Table:spatialaudioencoder} showcases the performance of our spatial audio encoder for seen and unseen audio classes. The DSAST model performs on par with the DCASE 2025 Task 3 baseline in F-score and relative distance error without using any synthetic training data. The slight decrease in F1 score for DSAST w/ AGM can be attributed to the model learning a joint audio-text embedding representation in addition to learning the spatial attributes. We expect this can be mitigated with extra training data to learn a robust audio-text alignment. For challenging scenario of unseen classes evaluation, our proposed model with AGM maintains strong generalization with an F-score of 0.327. It also achieves a lower DoA error of 25.05 while maintaining a reasonable relative distance error of 0.51. This experiments demonstrates that DSAST w/ AGM leverages the mono-audio event detection confidence from AGM and estimate their frame level spatial attributes, DoA and source distance, competitively and generalize well for unseen classes.

Table \ref{Table:Maintable} presents the end-to-end spatial audio question answering performance for the three models across various spatial audio reasoning capabilities on our spatial motion understanding and reasoning benchmark. The DSAST+Qwen7B model demonstrate strong and balanced performance across all capabilities, achieving the highest accuracy in Distance Dynamics (34.6\%), Cross-Source comparisons (30.0\%), and Dist vs DoA (33.3\%). On the other hand, DSAST+Qwen7B performs the best on DoA, achieving 35.8\% but struggles in other areas such as distance dynamics and cross source comparison. This could be due to error propagation from the class wise estimated spatial attributes to the reasoning LLM. This is observed more for complex question types such as cross source comparisons and distance vs DoA analysis. The low scores of BAT across all the categories indicates poor understanding of spatial information and reasoning on moving sources. This necessitates for an enhanced end-to-end training approach to better capture the fine-grained spatial understanding of audio events. The bar plot from Figure \ref{fig:barplotquestiontype} shows that boolean type questions reasoning is easier for all the models compared to the MCQ-Single questions. The DSAST w/ AGM + Qwen7B model maintains good performance on both the categories indicating better reasoning across different question formats.

\section{Conclusion}

In this work, we proposed a spatial audio encoder, DSAST w/ AGM, integrated with an audio grounding mechanism to localize and estimate spatial attributes of audio events across time frames. We further introduce an end-to-end training free spatial reasoning framework that leverages reasoning LLMs for structured prompting of the spatial attributes to answer complex queries. Additionally, we introduced a spatial motion understanding and reasoning benchmark to evaluate SAQA models on dynamic source movements. Our experiments demonstrate the effectiveness of the proposed model compared to baseline, while revealing that current SAQA models achieve only about 30\% accuracy on audio objects movement reasoning, highlighting the challenge of this task. Future work will focus on improving the framework, extending support to longer audio clips, and enabling efficient inference with smaller language models.




\bibliographystyle{IEEEbib}
\bibliography{strings,refs, IEEEexample, IEEEfull}

\end{document}